\documentclass[secnumarabic,aps,prl,reprint,groupedaddress]{revtex4-1}
\usepackage{verbatim}
\usepackage[english]{babel}
\usepackage{graphicx}
\usepackage{color}
\usepackage{amsmath}

\begin{document}

\title{Spin glass transition in a thin-film NiO/Permalloy bilayer}
\author{Tianyu Ma\thanks{Current affiliation: Department of Physics, Cornell University, Ithaca, NY 14853} and Sergei Urazhdin}
\affiliation{Department of Physics, Emory University, Atlanta, GA 30322}

\begin{abstract}

We experimentally study magnetization aging in a thin-film NiO/Permalloy bilayer. Aging characteristics are nearly independent of temperature below the exchange bias blocking temperature $T_B$, but rapidly vary above it. The dependence on the magnetic history qualitatively changes across $T_B$. The observed behaviors are consistent with the spin glass transition at $T_B$, with significant implications for magnetism and magnetoelectronic phenomena in antiferromagnet/ferromagnet bilayers.

\end{abstract}
\maketitle

The properties of thin-film antiferromagnets (AFs) interfaced with ferromagnets (Fs) first became the subject of intense research in the context of exchange bias (EB) - unidirectional anisotropy acquired by F when the system is cooled through a certain blocking temperature $T_B$~\cite{Meiklejohn56,NOGUES1999203}. They have recently attracted a renewed attention thanks to the high dynamical magnetization frequencies of AFs, enabling applications in fast interconversion and transmission of spin signals~\cite{PhysRevLett.118.067202}, and THz optics~\cite{Nishitani2010}. Additionally, vanishing magnetization of AFs can enable enhanced spin-transfer efficiency in electronic manipulation of the magnetic states for ultrahigh-density information storage, avoiding the constraints imposed by the angular momentum conservation and the dipolar fields ubiquitous to ferromagnetic systems~\cite{Jungwirth2016}.

A number of novel phenomena have been recently observed or predicted for thin AF films, including antiferromagnetic spin-orbit torques~\cite{PhysRevLett.113.196602,PhysRevB.92.165424,FukamiShunsuke2016Msbs}, AF magnetoresistance~\cite{SinovaJairo2012Nmot}, enhanced interconversion between electron spin current and spin waves~\cite{PhysRevLett.118.067202,PhysRevB.94.014427}, generation of THz signals~\cite{Nishitani2010,Khymyn2017}, AF exchange springs~\cite{PhysRevLett.92.247201,Hajiri2017}, and topological effects~\cite{PhysRevLett.87.116801,PhysRevB.95.035422,Hanke2017}. While some of these phenomena are expected even for standalone AFs, strong exchange coupling at AF/F interfaces provides one of the most efficient approaches to controlling and analyzing the magnetization states of AFs, with the state of F controlled by the magnetic field or spin current, and characterized by the magnetoelectronic or optical techniques. However, despite intense ongoing research, little is known about the dynamical and even static magnetization states of thin AF films in F/AF bilayers.

Here, we present measurements of magnetization aging, observed in a thin NiO film after magnetization reversal of an adjacent ferromagnet. Our main result is the observation of an abrupt transition between two qualitatively different aging regimes, which we identify as a glass transition in AF frustrated by the random exchange interaction at the F/AF interface. The insight provided by our findings may enable the implementation of new functionalities in magnetic nanodevices, facilitated by the controlled transition among the multidomain, spin-liquid, and spin-glass states of thin-film magnetic systems.

The AF in our study was a polycrystalline $15$~nm-thick NiO layer deposited by reactive sputtering on an oxidized  $6\times 6$~mm$^2$ Si substrate. A $10$-nm-thick Ni$_{80}$Fe$_{20}$=Permalloy (Py) ferromagnet and a $3$~nm-thick capping SiO$_2$ layer were deposited on top. NiO has played a prominent role in recent studies of magnetic, magnetoelectronic, and optical phenomena in AFs, thanks to its simple electronic, magnetic, and crystalline structure~\cite{PhysRevLett.92.247201,PhysRevLett.118.067202,PhysRevB.85.174439,Nishitani2010,PhysRevB.94.014427}. Additionally, the magnetocrystalline anisotropy of NiO is more than an order of magnitude smaller than that of other common AF such as CoO or Fe$_{50}$Mn$_{50}$~\cite{BERKOWITZ1999552}, making NiO promising for the manipulation of its magnetization. These characteristics facilitated our observations, as discussed below.

To characterize the state of the studied system, we utilized the anisotropic magnetoresistance (AMR) of Py, measured in the van der Pauw geometry. The field $H$ was applied in-plane perpendicular to the average direction of current. In this configuration, the AMR is described by $R(\theta)=R_{min}+\Delta R sin^2(\theta)$, where $\theta$ is the average angle formed by the magnetization $\mathbf{M_F}$ of Py relative to $\mathbf{H}$, $R_{min}$ is the AMR minimum at $\theta=0$, and $\Delta R$ is the magnetoresistance. Since the magnetic anisotropy of Py is negligible on the scale of typical fields in our measurements, the temporal evolution of $R$ observed at fixed experimental conditions was caused by the magnetic aging of NiO, resulting in the variations of the effective exchange field exerted by NiO on Py.

\begin{figure}
\includegraphics[width=1\columnwidth]{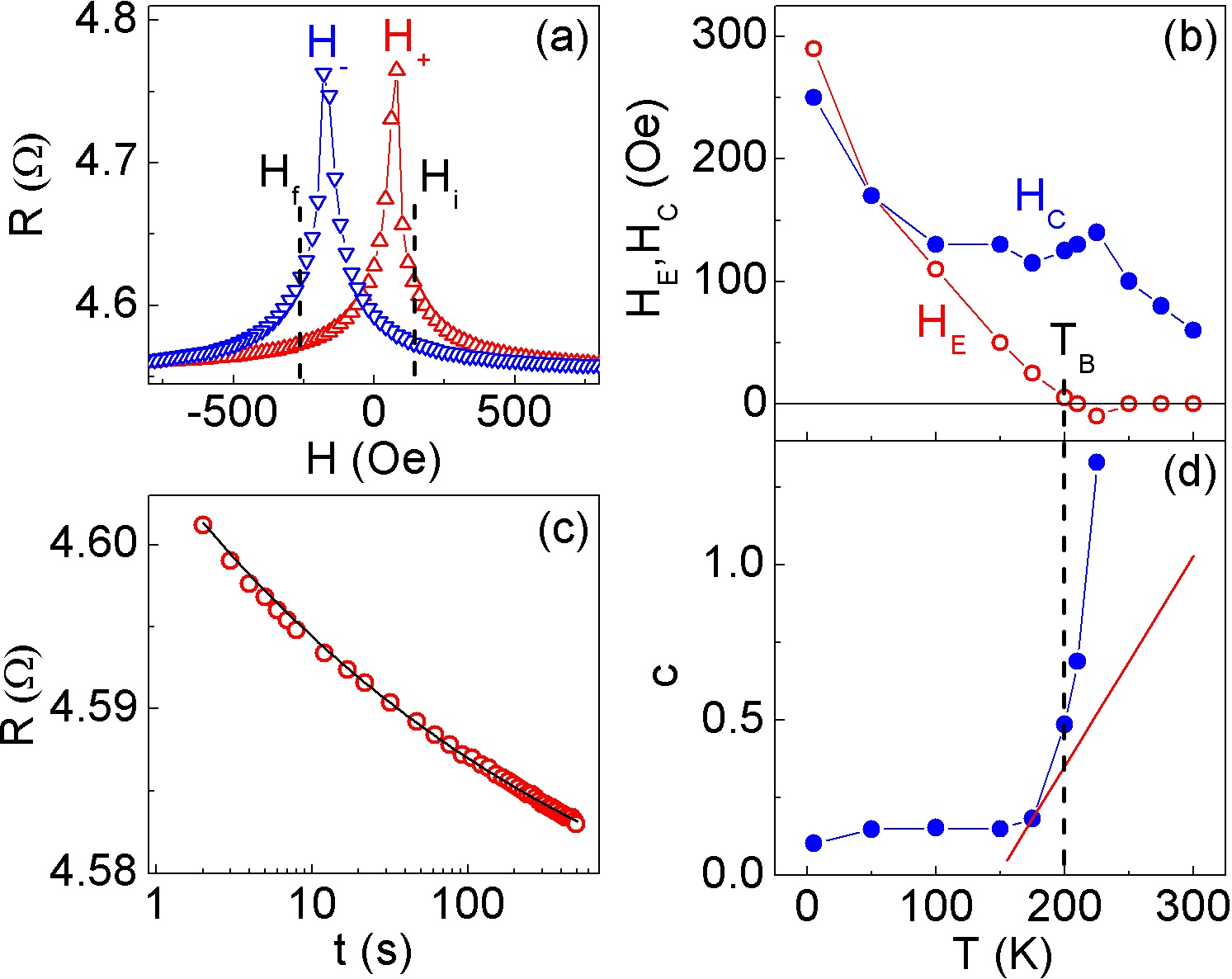} 
\caption{\label{fig:1} (a) Magnetoelectronic hysteresis loop obtained at $T=150$~K after two prior similar "training" loops. Up and down triangles are for the increasing- and decreasing-field sweeps, respectively. The coercive fields $H_-$ and $H_+$ are labeled. Vertical dashed lines, labeled $H_i$ and $H_f$, indicate the field values used in the measurements of aging. (b) Temperature dependence of coercivity $H_{C}$ and of the exchange bias field $H_{E}$. (c) Symbols: time evolution of resistance $R$ at field $H_f=-260$~Oe, after pre-aging at field $H_i=140$~Oe. Curve is the result of fitting with the power-law dependence $R(t) = R_0 + At^{-c}$, with the best-fit value $c=0.15$. Note the logarithmic time scale. (d) Temperature dependence of exponent $c$, obtained by fitting the aging data as shown in panel (c). Curve: temperature dependence of $c$ calculated based on the Arrhenius model, with the distribution of activation barriers determined from the data for $T=175$~K. The vertical dashed line through (b) and (d) shows the blocking temperature $T_B$.}
\end{figure}

In the scans of $H$, the resistance exhibited sharp peaks at the coercive fields $H_{-}$ and $H_{+}$ [Fig.~\ref{fig:1}(a)]. The exchange bias field $H_E=(H_{+} + H_{-})/2$ and the coercivity $H_c=(H_{+} - H_{-})/2$ decrease with increasing $T$, Fig.~\ref{fig:1}(b). The exchange bias vanishes at the blocking temperature $T_B=200$~K, while the coercivity exhibits a peak at temperatures slightly above $T_B$, consistent with the prior studies of NiO/F bilayers~\cite{PhysRevB.58.8566}.

To study magnetic aging, we first "pre-aged" the system by applying a field $H_i>H_{+}$ over the time interval $\Delta t_i$ set to $500$~s in measurements described below, unless specified otherwise. The field was then reversed to $H_f<H_{-}$, and the resistance was subsequently recorded in $1$~s time steps for $500$~s. The values of $H_i$ and $H_f$ were chosen so that the corresponding values of $R$ in the hysteresis loop were larger than $R_{min}$ by approximately $0.3$ of the full MR. The resulting average angle between $\mathbf{M_F}$ and $\mathbf{H}$ was $\theta\approx30^\circ$. Under these conditions, the measured resistance variations due to aging were approximately proportional to $\theta$, which is in turn approximately proportional to the transverse to $\mathbf{H}$ component of the effective exchange field exerted by NiO on Py. Thus, the time dependence of $R$ provided a direct quantitative measure of magnetization aging in NiO~\cite{Ma2016}.

The aging data were well-approximated by the power-law time dependence $R=R_0+At^{-c}$~\cite{Urazhdin2015,Ma2016}, with the asymptotic resistance $R_0$, amplitude $A$, and the aging exponent $c$ used as the fitting parameters [see Fig.~\ref{fig:1}(d) for $T=150$~K]. At $T<T_B=200$~K, the value of $c$ remained small, increasing from $0.10$ at $T=5$~K to $0.18$ at $T=175$~K [Fig.~\ref{fig:1}(d)]. At $T>T_B$, it started to rapidly increase, reaching $c=1.33$ at $T=225$~K. At $T>225$~K, aging became too rapid for reliable measurements with our technique, but its signatures persisted in the aging data.

Our main new result is the observation of an abrupt transition between two qualitatively different aging regimes at $T_B$. To highlight its fundamental significance, we consider the Neel-Brown-Arrhenius theory commonly utilized to describe thermal dynamics of thin polycrystalline AF films in F/AF bilayers~\cite{FulcomerCharap72,PhysRevB.59.3722,OGrady2010}. The distribution of the activation barriers of the AF grains can be determined from the aging data at a given temperature $T_0$, allowing one to calculate the expected temperature dependence of aging characteristics. For the power-law aging, this model predicts $c(T)=[(1+c(T_0))T/T_0-1]$~\cite{Ma2016}. Here, we neglect the dependence of the activation barriers on $T$ far below the Neel temperature of AF. The Arrhenius dependence, calculated using $T_0=175$~K and $c(T_0)=0.18$ [solid line in Fig.~\ref{fig:1}(d)], falls below the data both at $T<T_B$ and $T>T_B$. Thus, the activation is sub-Arrhenius at $T<T_B$, and super-Arrhenius at $T>T_B$. 

The super-Arrhenius temperature dependence at $T>T_B$ is reminiscent of the super-cooled liquid state of glass-forming systems above the glass transition~\cite{Ediger1996,Mirigian2013}. Meanwhile, the slow, almost temperature-independent aging at $T<T_B$ is similar to the avalanche dynamics in fragile systems such as sandpiles~\cite{PhysRevLett.81.1841,Kardar1996}. 
To elucidate these similarities, we note that the magnetization of AF in F/AF bilayers experiences a random effective exchange field ubiquitous to F/AF interfaces due to their unavoidable roughness~\cite{Malozemoff1987,Malozemoff1988}. For F/AF alloys, frustration caused by the random exchange fields of similar origin can lead to the formation of a spin glass~\cite{RevModPhys.58.801}. In F/AF bilayers, the random effective exchange field experienced by AF scales inversely with the AF thickness $d$. Below a certain thickness $d_{crit}$, it was predicted to produce a new spin state termed by Malozemoff the Heisenberg domain state~\cite{Malozemoff1988}. This state is distinct from the multidomain (Imry-Ma) state in thicker AF films~\cite{Ma1975,Malozemoff1987,PhysRevB.66.014430}, because the magnetic anisotropy stabilizing the AF domains is negligible compared to the random field, so the only vestige of AF ordering remaining in this state is the local spin correlation. Similar local structural correlations exist in amorphous solids~\cite{Elliott1991}. We identify this state as a spin glass, as proposed in some models of exchange bias~\cite{PhysRevB.38.6847,Kiwi1999,Radu2005}, and the transition at $T_B$ as the glass transition. We provide further experimental evidence for this interpretation below.

The observed transition is likely generic to thin-film F/AF systems. Indeed, slow power-law aging was observed for in F/AF bilayers based on AF=Fe$_{50}$Mn$_{50}$ and CoO with $d\le 4$~nm~\cite{Urazhdin2015, Ma2016}. The smaller values of $d$ needed to observe aging in these AFs can be attributed to their much larger anisotropy constant $K$~\cite{BERKOWITZ1999552}, consistent with $d_{crit}\propto 1/\sqrt{K}$ predicted by Malozemoff~\cite{Malozemoff1988}. For such ultrathin AF films, local variations of thickness due to roughness become comparable to $d$, resulting in large local variations of the transition temperature. A spatially inhomogeneous mixture of spin-glass and spin-liquid - the spin-slush - can then form at temperatures around $T_B$. The resulting inhomogeneous broadening obscures the glass transition.

\begin{figure}
\includegraphics[width=1\columnwidth]{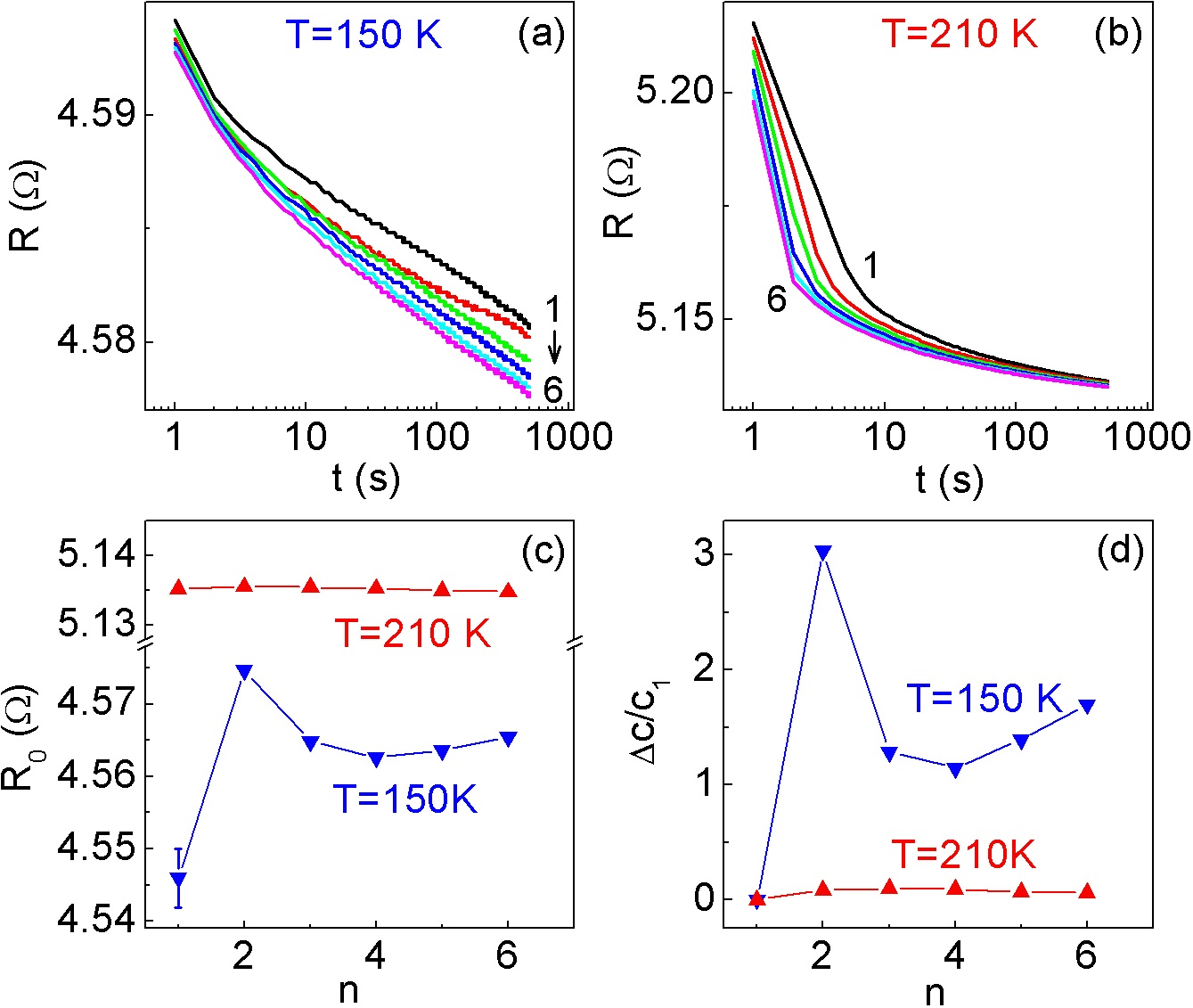} 
\caption{\label{fig:2} (a) Sequential aging measurements performed at $H_f=-260$~Oe after pre-aging at $H_i=140$~Oe, immediately after cooling to $T=150$~K. (b) Same as (a), at $T=210$~K, $H_f=-150$~Oe, $H_i=150$~Oe.  (c,d) Dependence of the asymptotic resistance $R_0$ (c) and of the normalized variation $\Delta c/c_1=(c_n-c_1)/c_1$ of the power-law exponent (d) on the aging cycle number $n$. The fitting errors were smaller than the symbol sizes, except for $n=1$ in (c), as indicated by an error bar.} 
\end{figure}

To test our interpretation of the transition observed at $T_B$, we analyzed the dependence of aging on the magnetic history. The spin glass state at $T<T_B$ should be irreversibly perturbed by the reversal of $\mathbf{M_F}$, resulting in different asymptotic Gibbs states in each sequential aging measurement~\cite{RevModPhys.58.801}. In contrast,  the spin-liquid at $T>T_B$ should asymptotically reach the same equilibrium state, regardless of history~\cite{Ediger1996,Mirigian2013}. Indeed, the aging curves, measured at $T=150$~K$<T_B$ immediately after cooling, noticeably vary with the aging cycle [Fig.~\ref{fig:2}(a)]. At $T=210$~K$>T_B$, the difference among the aging curves is significant only in the first few seconds [ Fig.~\ref{fig:2}(b)], which can be attributed to minor variations of the initial state in the fast aging process. These conclusions are supported by the quantitative analysis. At $T=150$~K, both the asymptotic resistance $R_0$ [Fig.~\ref{fig:2}(c)], and the normalized power-law exponent variations $\Delta c/c_1=(c_n-c_1)/c_1$  [Fig.~\ref{fig:2}(c)], jump between $n=1$ and $2$, and continue to vary in subsequent aging cycles. These variations indicate that the system asymptotically approaches different Gibbs states in each aging cycle, with the initial "annealed" state being substantially different from the local minima of the energy landscape reached in subsequent aging measurements. In contrast, at $T=210$~K, the variations of $R_0$ are negligible, while $\Delta c/c_1$ remains small, indicating that all the aging cycles are statistically equivalent. This is consistent with the lack of long-time temporal correlations in the spin-liquid state~\cite{Ediger1996,Mirigian2013}.

\begin{figure}
\includegraphics[width=1\columnwidth]{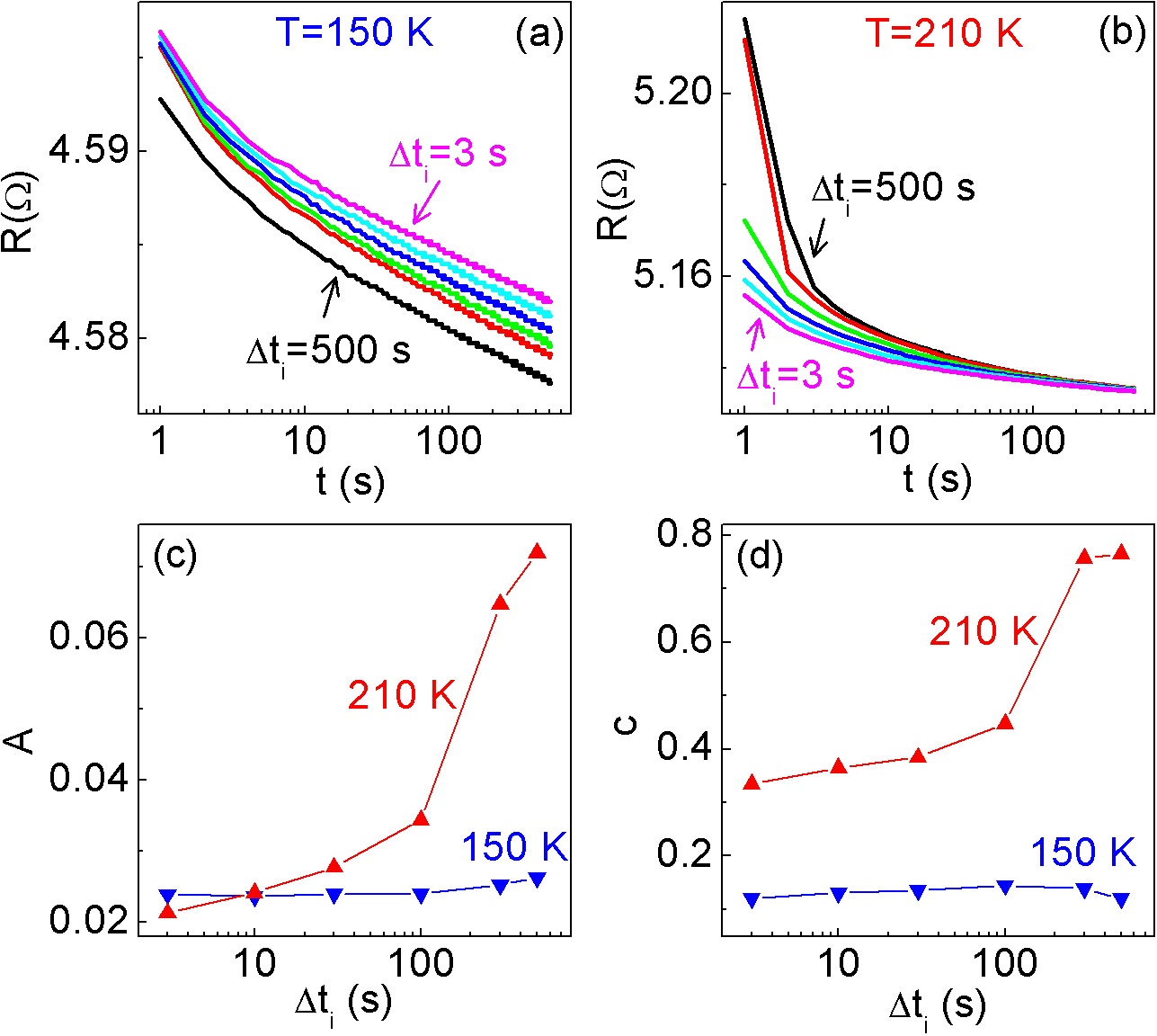} 
\caption{\label{fig:3} (a,b) R vs t at $T=150$~K (a) and $T=210$~K (b) after prior aging over time interval $\Delta t_i =3$~s, $10$~s, $30$~s, $100$~s, $300$~s, and $500$~s. (c,d) Dependence of relaxation scale A (c), and power-law exponent c on $\Delta t_i$. Point-down (point-up) triangles are for $T=150$~K ($210$~K). The fitting errors in (c) and (d) are smaller than the symbol sizes, and are not shown for clarity.}
\end{figure}

The qualitative difference between the states of the system below and above $T_B$ was elucidated by the dependence of aging on $\Delta t_i$, Fig.~\ref{fig:3}, which also unequivocally demonstrated that Arrhenius relaxation is irrelevant for the studied system. We start by analyzing the effects of $\Delta t_i$ on the Arrhenius aging. We assume that the system can be described by some distribution of activation barriers, and aging at $\Delta t_i\to\infty$ is described by the power law with exponent $c$. For finite $\Delta t_i$, the Arrhenius model then predicts power-law aging with exponent $(c+1)$ at $t>\Delta t_i$~\cite{Ma2016}. The increase of the exponent reflects a larger relative contribution of the fast activation, since the slow degrees of freedom do not become activated during short pre-aging. 

Figures~\ref{fig:3}(a) and (b) show the aging curves for $\Delta t_{i}$ ranging from $3$~s to $500$~s, at $T=150$~K and $210$~K, respectively. The measurements were performed after at least six prior aging cycles at the same temperature. At $T=150$~K, the aging curves slightly shifted when $\Delta t_i$ was varied, but their shape was independent of $\Delta t_{i}$. This is consistent with the cooperative nature of avalanche dynamics, which cannot be characterized by any distribution of characteristic activation timescales~\cite{PhysRevLett.59.381,Kardar1996}. In contrast, at $T=210$~K, the aging curves exhibit a strong dependence on $\Delta t_i$. Most notably, the variations of $R$ over the first few seconds of aging become dramatically enhanced at $\Delta t_i > 100$~s. Quantitative analysis supports these observations. At $T=150$~K, both the aging amplitude $A$ [Fig.~\ref{fig:3}(c)] and the power-law exponent $c$ [Fig.~\ref{fig:3}(d)] are almost independent of $\Delta t_i$. Meanwhile, at $T=210$~K, both $A$ and $c$ significantly increase with increasing $\Delta t_i$. The dependence $c(\Delta t_i)$ observed at $T=210$~K  {\it qualitatively disagrees with the Arrhenius prediction}: in our measurements, the value of $c$ decreases at small $\Delta t_i$, instead of the increase predicted for the Arrhenius activation.

\begin{figure}
	\includegraphics[width=1\columnwidth]{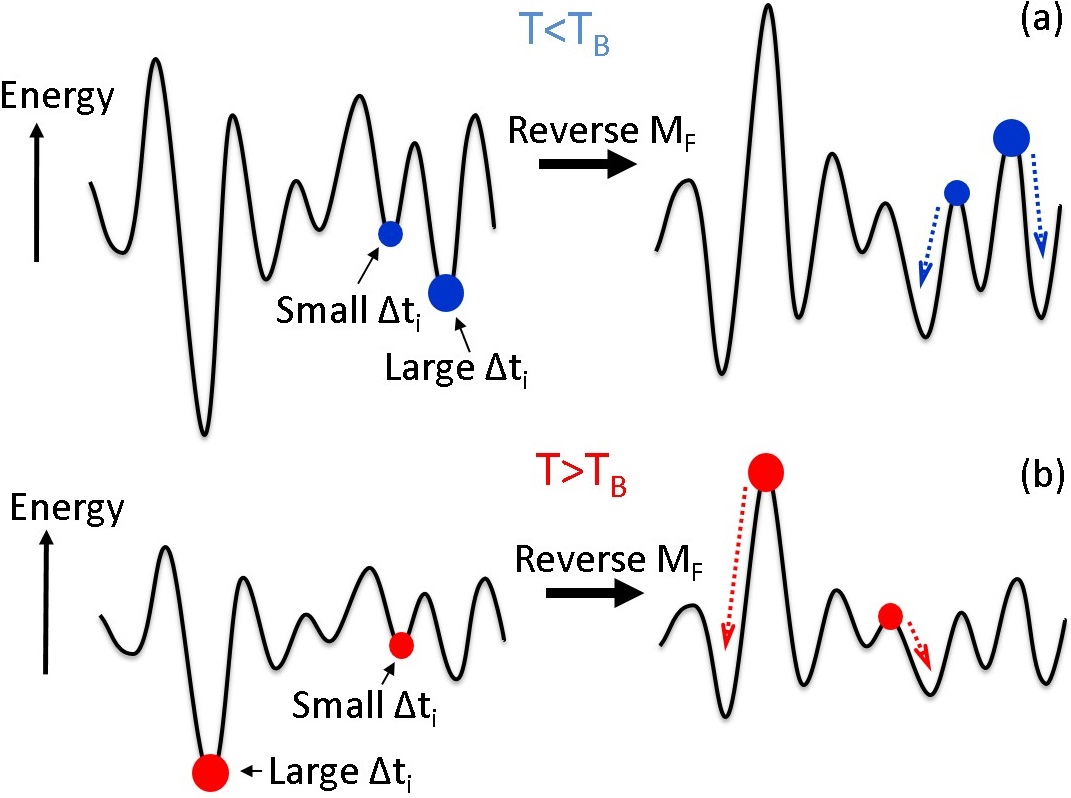} 
	\caption{\label{fig:4} Dependence of aging on the pre-aging duration $\Delta t_i$. The energy landscape is schematically shown in the spin-glass (a) and the super-cooled liquid states (b), after pre-aging (left) and immediately after the reversal of $\mathbf{M_F}$ (right). Small and large filled circles indicate the state of the system after short and long pre-aging, respectively. Dotted arrows indicate the evolution of the system after the reversal.}
\end{figure}

To interpret our observations, we consider the energy landscape of the frustrated spin system, Fig.~\ref{fig:4}. At $T<T_B$ in the spin-glass state, the largest activation barriers significantly exceed thermal energy, so that only a small volume of the configuration space is accessible via thermal activation. After pre-aging, the system becomes trapped in a shallow energy minimum both for small and large $\Delta t_i$ [Fig.~\ref{fig:4}(a), left]. The energy landscape, determined by the exchange interaction across the F/AF interface, is reversed when $\mathbf{M_F}$ is reversed by the field $H_f$. The resulting evolution of the system, starting from a shallow energy maximum, is nearly independent of $\Delta t_i$ [dotted arrows in Fig.~\ref{fig:4}(a), right].

In the supercooled spin-liquid state at $T>T_B$, the largest activation barriers are comparable to thermal energy, so the whole configuration space becomes accessible~\cite{Ediger1996,Mirigian2013}. After short pre-aging, the system is in a shallow energy minimum, but at large $\Delta t_i$ it reaches a deep minimum [Fig.~\ref{fig:4}(a), left]. After the reversal of $\mathbf{M_F}$, this minimum becomes a sharp maximum, resulting in more rapid initial aging [Fig.~\ref{fig:4}(b), right]. This analysis explains not only the observed increase of $c$ at large $\Delta t_i$, but also the large difference between the timescales of pre-aging and of its effects on aging: variations of the former in the $100$~s range mostly affect the first few  seconds of subsequent aging [see Fig.~\ref{fig:3}(b)]. Indeed, the initial rate of relaxation is determined by the energy profile of the deep energy minimum, while the time to reach this minimum (the characteristic pre-aging time) is determined by the diffusion rate through the configuration space.

To summarize, measurements of aging in a thin-film NiO/Permalloy bilayer revealed the hallmark features of a spin-glass transition at the exchange bias blocking temperature $T_B$. At $T<T_B$, the Gibbs state of the system depends on its magnetic history, and its aging characteristics are consistent with the avalanche spin dynamics expected for a fragile spin-solid. At $T>T_B$, the system relaxes to the same equilibrium state, independent of history. The large effects of magnetic history on the aging dynamics in this state qualitatively disagree with the Arrhenius activation, but are consistent with the short-timescale correlations in a supercooled spin liquid. 

A number of significant implications of these findings are expected for the magnetic and magnetoelectronic phenomena in F/AF bilayers.  The demonstrated new type of spin glass may provide insight into the general mechanisms underlying the formation of ferroic glasses~\cite{Sherrington2014}. For the applications of NiO and other ultrathin AF films in spin-charge conversion and magnonic (spin wave-based) structures~\cite{PhysRevLett.87.116801,Jungwirth2016,PhysRevLett.118.067202,PhysRevB.94.014427}, we expect a transition at $T_B$ between two different regimes of spin wave propagation, and a large difference between the spin wave characteristics in the multidomain state of thick AF films and the spin-glass state of thin films. Surprisingly, a recent study suggests that spin disorder may result in enhanced spin-wave propagation~\cite{WesenbergDevin2017Lsti}.  Furthermore, magnetic frustration in AF can relax the spin-wave momentum conservation, resulting in enhanced spin interconversion at AF interfaces~\cite{PhysRevLett.118.067202}. The formation of the AF exchange springs~\cite{PhysRevLett.92.247201} and the nanoscale writing of exchange bias~\cite{AlbisettiE2016Nrml} are governed by the  dynamical characteristics of the F/AF system~\cite{Venus2005}, which can be controlled by engineering the transition between the spin-liquid and spin-solid states. Frustration governing the observed aging phenomena may be also important for the topological properties of the AF spin state~\cite{PhysRevLett.118.247203,Proctor2014}.

Our analysis also showed that the spin glass transition is controlled not only by the thickness, but also by the anisotropy of the AF~\cite{Hajiri2017}. Based on this observation, we predict a new multiferroic effect for F/AF bilayers~\cite{WuSM2010Reco}, whereby variations of the magnetic anisotropy, due either to the direct effects of electric field or to magnetostriction in a hybrid piezoelectric/F/AF heterostructure, can produce electronic switching, at a fixed temperature, between the spin-solid and spin-liquid states. This can provide an efficient and robust method to electronically control the static and dynamical characteristics of magnetic nanostructures.

\begin{acknowledgements}
	We acknowledge support from the NSF Grant DMR-1504449.
\end{acknowledgements}
\bibliography{NiO}{}
\bibliographystyle{apsrev4-1}

\end{document}